\documentclass[letterpaper, 10 pt, conference]{ieeeconf}  

\IEEEoverridecommandlockouts                              

\usepackage{graphics} 
\usepackage{times} 
\usepackage{amsmath} 
\usepackage{amssymb}  
\allowdisplaybreaks
\usepackage{balance}

\title{\LARGE \bf
Distributed Traffic Signal Control via Coordinated Maximum Pressure-plus-Penalty 
}

\author{Vinzenz T\"utsch$^{1}$, Zhiyu He$^{2}$, Florian D\"orfler$^{2}$, and Kenan Zhang$^{1}$
\thanks{$^{1}$ Human-oriented mobility eco-system (HOMES) laboratory, EPFL, {\tt\small \{vinzenz.tutsch, kenan.zhang\}@epfl.ch}}
\thanks{$^{2}$ Automatic Control Laboratory (IfA), ETH Z\"urich, {\tt\small \{zhiyhe, dorfler\}@control.ee.ethz.ch}}
}

\newcommand{\setG}{\mathcal{G}}
\newcommand{\setI}{\mathcal{I}}
\newcommand{\setN}{\mathcal{N}}
\newcommand{\setM}{\mathcal{M}}
\newcommand{\setS}{\mathcal{S}}
\newcommand{\setL}{\mathcal{L}}
\newcommand{\setLentry}{\mathcal{L}_\textup{entry}}
\newcommand{\setLexit}{\mathcal{L}_\textup{exit}}
\newcommand{\setLinternal}{\mathcal{L}_\textup{internal}}
\newcommand{\setLi}{\mathcal{L}_{i}}
\newcommand{\setD}{\mathcal{D}}
\newcommand{\setU}{\mathcal{U}}
\newcommand{\bbE}{\mathbb{E}}
\newcommand{\boldQ}{\mathbf{Q}}
\newcommand{\boldd}{\mathbf{d}}
\newcommand{\boldx}{\mathbf{x}}
\newcommand{\boldz}{\mathbf{z}}
\newcommand{\boldlambda}{\boldsymbol{\lambda}}

\usepackage{graphicx} 
\usepackage{url}
\usepackage{caption}
\usepackage{cleveref}
\usepackage[per-mode=symbol]{siunitx}
\usepackage{algorithm}
\usepackage{algpseudocode}
\usepackage{subcaption}

\usepackage{booktabs}

\newtheorem{theorem}{Theorem}

\newtheorem{definition}{Definition}
\newtheorem{assumption}{Assumption}

\makeatletter
\def\algbackskip{\hskip-\ALG@thistlm}
\makeatother

\usepackage{color}

\begin{document}

\maketitle
\thispagestyle{empty}
\pagestyle{empty}

\begin{abstract}
This paper develops an adaptive traffic control policy inspired by Maximum Pressure (MP) while imposing coordination across intersections. The proposed Coordinated Maximum Pressure-plus-Penalty (CMPP) control policy features a local objective for each intersection that consists of the total pressure within the neighborhood and a penalty accounting for the queue capacities and continuous green time for certain movements. 
The corresponding control task is reformulated as a distributed optimization problem and solved via two customized algorithms: one based on the alternating direction method of multipliers (ADMM) and the other follows a greedy heuristic augmented with a majority vote. 
CMPP not only provides a theoretical guarantee of queuing network stability but also outperforms several benchmark controllers in simulations on a large-scale real traffic network with lower average travel and waiting time per vehicle, as well as less network congestion. Furthermore, CPMM with the greedy algorithm enjoys comparable computational efficiency as fully decentralized controllers without significantly compromising the control performance, which highlights its great potential for real-world deployment.  





\end{abstract}


\section{Introduction}
\subsection{Background and Motivations}\label{sec:background}
Vehicular traffic in urban areas has surged dramatically over the past decades~\cite{czepkiewicz_Why_2018} and is expected to continue growing in future~\cite{gov.uk_National_}. This forecast poses significant challenges to traffic control~\cite{jayasooriya_Measuring_2017a}.
In dense urban road networks, traffic signals have long constituted effective control instruments to regulate traffic flows and ensure efficient vehicle movements. 
In the literature, Traffic Signal Controllers (TSCs) are often classified into three types: fixed-time, actuated, and adaptive. 
Specifically, fixed-time TSCs are the most commonly used in practice, which allocate a fixed green time to each movement~\cite{barman_performance_2022}. The actuated TSCs set green time in response to real-time traffic volumes, though the control policy is predefined. Differently, adaptive TSCs optimize the control actions to maximize the traffic throughput. 
Early developments of adaptive signal control relied on centralized approaches (e.g., \cite{hunt_scoot_1982,lowrie_scats_1990,henry_prodyn_1983}) and thus can hardly be implemented in real practice due to rapidly increasing computational and communication costs in large-scale traffic networks. 

The scalability issue of classic adaptive TSCs was addressed by the Maximum Pressure (MP) control, a decentralized algorithm proposed by \cite{varaiya_Max_2013}. In short, at each time, MP selects a signal phase that maximizes its ``pressure'', a measure computed solely using the number of vehicles on incoming and outgoing lanes associated with the phase. Accordingly, MP can be solved efficiently without communication among interactions or central coordination. 
In addition to satisfactory performances~\cite{sun_simulation_2018, lioris_adaptive_2016}, MP is widely celebrated for its strong theoretical underpinning. Using the Lyapunov drift theory, \cite{varaiya_Max_2013} proved that MP guarantees the stability of the stochastic queuing processes at all intersections. 
However, a key assumption for the proof is infinite queue capacity, which may not hold for dense road networks. To address this limitation, several recent studies have modified the original MP algorithm to explicitly consider limited queue lengths \cite{gregoire_CapacityAware_2015,xiao_pressure_2014}. 
Another issue with MP is the possibly extensive time of red lights for certain phases. To remedy this, cyclic algorithms have been introduced that ensure each phase is activated at least once within a period \cite{le_Decentralized_2014,levin_Maxpressure_2020}.
All these MP variants remain fully decentralized and are shown to maintain the stability property under certain conditions. 

Recently, several studies leveraged the Lyapunov minimum drift-plus-penalty control~\cite{neely_Stochastic_2010}, an extension of the theory that MP is built upon, to tackle the issue of finite queue lengths. 
In \cite{bracciale_Lyapunov_2020}, a finite queue length is enforced as a constraint, while a queue-length-dependent penalty is appended to the regular pressure for each intersection in \cite{hao_Backpressure_2020}. 
Although these algorithms leverage information from neighboring intersections, they still optimize control actions for each agent independently, without explicit coordination. These decentralized approaches may lead to suboptimal performance because the individual intersections do not account for broader network effects and interdependencies.

To augment the MP framework with coordination and attention to queue capacity constraints, this paper reformulates the network signal control problem into a distributed optimization~\cite{yang_Survey_2019}, while exploiting the Lyapunov minimum drift-plus-penalty control to define local objectives and to prove the network queuing stability. Specifically, each intersection aims to maximize the total pressure within its neighbor, penalized by its impact on neighboring intersections. 
Consequently, the coordinated signal control policy can be solved in an online and adaptive manner as per other MP methods (e.g., \cite{varaiya_Max_2013,gregoire_CapacityAware_2015}) while achieving network-wide global optimum. 


\subsection{Contributions}
This paper develops a distributed traffic signal control policy, namely, Coordinated Maximum Pressure-plus-Penalty (CMPP), which extends the standard MP approach by leveraging communication and coordination across neighboring intersections. 
Our key contributions include:

\begin{itemize}
    \item Expand the per-intersection pressure to a neighborhood of intersections and augment it with a penalty that captures queue capacities and continuous green time of certain vehicle movements.

    \item Establish the stability of the queuing network under the CMPP control policy using the Lyapunov optimization theorem~\cite{neely_Stochastic_2010}. 
    
    \item Reformulate the network control problem as distributed optimization and develop two consensus algorithms based on the alternating direction method of multipliers (ADMM) and a greedy heuristic.

    \item Demonstrate the performance of CMPP through simulations of real-world road networks and traffic demand.
\end{itemize}


\section{Problem Setting and Preliminaries}
\label{sec:problem_and_preliminaries}
\subsection{Road Network, Signal Control, and Queue Dynamics}
Consider a directed road network $\setG = (\setL, \setI)$, where $\setI$ denotes the set of intersections with $|\setI|=N$ and $\setL$ denotes the set of road links. We further divide $\setL$ into three subsets: i) entry links $\setLentry$, where vehicles enter the network; ii) internal links $\setLinternal$ that connect intersections inside the network; and iii) exit links $\setLexit$, from which vehicles leave the network.

\begin{figure}
    \centering
    \includegraphics[width=0.4\textwidth]{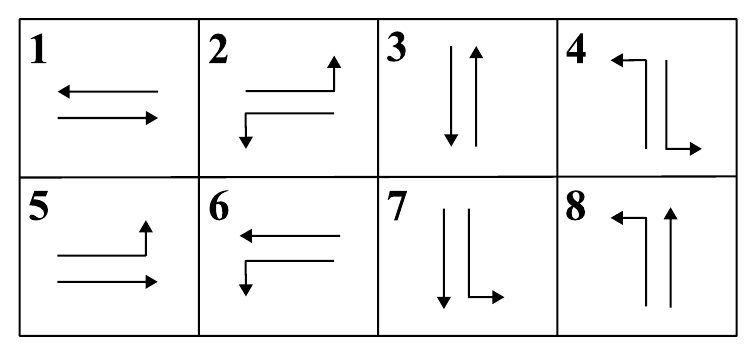}
    \caption{Eight phases in a typical intersection.}
    \label{fig:phases}
\end{figure}

Each intersection features a set of links $\setLi\subset \setL$ that denote the traffic movements through it. 
The tuple $(l,m)$ with $l,m\in\setLi$ defines the movement from link $l$ to link $m$ in intersection $i$.
Accordingly, the traffic signal control corresponding to each movement is given by $s_{l,m}\in\{0,1\}$. Specifically, $s_{l,m} = 1$ indicates that the movement $(l,m)$ is activated (i.e., a green light is on) and vice versa. 
Since multiple movements can be activated simultaneously without collision, the signal control action is always defined on \emph{phase}, a combination of movements. In this paper, we consider a configuration of eight typical phases\footnote{In this paper, we consistently allocate green time to right turns.} as depicted in \Cref{fig:phases} and use $\Phi_i$ to denote the set of phases for intersection $i$ with size $K_i=|\Phi_i|$. 
Let $\phi_{i,k}\in\{0,1\}$ denote the phase control of $k\in\Phi_i$. Then, $\phi_{i,k}=1$ implies $s_{l,m}=1,\forall (l,m)\in k$. 

Suppose each movement uses dedicated lanes. Then, the queue length corresponding to movement $(l,m)$ at time $t$, denoted by $q_{l,m}(t)$, can be modeled as follows:
\begin{align}
    \begin{split}
        q_{l,m}(t+1) &= q_{l,m}(t) - y_{l,m}(t) s_{l,m}(t) \\
        &\quad+ \left(\sum_{k \in \setU_l} y_{k,l}(t) s_{k,l}(t) + d_l(t) \right) r_{l,m}(t),
    \end{split}  \label{eq:store_and_forward_1}
\end{align}
where $y_{l,m}(t)$ gives the vehicle outflows 
and equals the minimum between the current queue length $q_{l,m}(t)$ and the capacity of the movement $c_{l,m}(t)$ (the maximum vehicle flow that can pass the intersection over single green time), i.e., 
\begin{align}
    y_{l,m}(t) &= \min\{q_{l,m}(t), c_{l,m}(t)\}.\label{eq:store_and_forward_2}
\end{align}
Accordingly, $y_{l,m}(t) s_{l,m}(t)$ gives the realized outflow of under the signal control at time $t$. 
On the other hand, the inflow is determined by the third term in \eqref{eq:store_and_forward_1}, where $\setU_l$ denotes the set of upstream links of link $l$, $d_l(t)$ represents the demand entering the network from link $l$, and $r_{l,m}(t)$ is the turning ratio as the proportion of vehicles on link $l$ moving to link $m$. Note that $d_l(t)$ is only non-zero for entry links $l \in \setLentry$ with $\setU_l=\emptyset$.

\subsection{Queuing Stability and Lyapunov Drift Theory} \label{sec:Queuing_stability_and_Lyapunov_drift_theory}

The queuing network stability represented by the vector $\boldQ(t)=[\dots,q_{l,m}(t),\dots]\in \mathbb{R}_+^{|\Lambda|}$, where $\Lambda$ denotes the set of all movements in the network, is defined as follows.

\begin{definition}[Stability of queuing process] \label{def:stability}
    The stochastic queuing process $\boldQ(t)$ is strongly stable if
    \begin{equation}
      \underset{t \to \infty}{\lim \sup} \frac{1}{t} \sum_{\tau=0}^{t-1} \mathbb{E}\{|\boldQ(\tau)|\} < \infty.
    \end{equation}
\end{definition}

A well-known approach to proving queue stability is the Lyapunov drift theory (e.g., \cite{varaiya_Max_2013,bracciale_Lyapunov_2020}). Specifically, a stability condition is constructed with a conditional Lyapunov drift defined as
\begin{equation} \label{eq:Lyapunov_drift}
    \Delta(\boldQ(t)) \overset{\triangle}{=} \bbE \left\{ L(\boldQ(t + 1)) -  (\boldQ(t)) \; \vert \; \boldQ(t) \right\},
\end{equation}
where $L(\boldQ(t)$ denotes the Lyapunov function, and a commonly used one is 
\begin{equation} \label{eq:Lyapunov_function}
    L(\boldQ(t)) \overset{\triangle}{=} \frac{1}{2} \sum_{(l,m)\in \Lambda} q_{l,m}(t)^2.
\end{equation}
The Lyapunov drift theory~\cite{neely_Stochastic_2010} states that if the initial queues satisfy $\bbE\{L(\boldQ(0))\} < \infty$, then a control policy fulfilling the following condition guarantees the strong stability of $\boldQ(t))$:
\begin{equation} \label{eq:Lyapunov_stability_inequality}
    \Delta(\boldQ(t)) \leq B - \epsilon
    \sum_{(l,m)\in\Lambda} |q_{l,m}(t)|,
\end{equation}
for some constants $B \geq 0, \epsilon > 0$.

\subsection{Maximum Pressure Control}
\label{sec:MP}

Maximum Pressure (MP) is a feedback control law that selects the active phase solely based on local queue lengths~\cite{varaiya_Max_2013}. At each time step $t$, a weight variable $w_{l,m}(t)$ is first computed for each movement $(l,m)$ as
\begin{equation*}
    w_{l,m}(t) = q_{l,m}(t) - \sum_{p \in \setD_m} r_{m,p}(t) q_{m,p}(t),
\end{equation*}
where $\setD_m$ denotes the set of downstream links from link $m$. These weights are then used to compute the \emph{pressure} of each phase $k$ per intersection $i$ as follows:
\begin{align*}
    \gamma_{i, k}(t) = \sum_{(l,m) \in k} c_{l,m}(t) w_{l,m}(t) s_{l,m}(t).
\end{align*}
The MP controller then activates the phase with the maximal pressure. Let $\phi^\text{MP}_i(t)$ be the MP control at intersection $i$ at time $t$. Then,
\begin{equation} \label{eq:MP_policy}
    \phi^\text{MP}_{i,k}(t + 1) = 
    \begin{cases}
        1, & k = \arg\max_{k'\in\Phi}\{ \gamma_{i,k'}(t)\},\\
        0, & \text{otherwise}. 
    \end{cases}
\end{equation}
Note that MP policy \eqref{eq:MP_policy} can be computed independently at each intersection and thus is fully decentralized.
It has also been proven that the MP policy is equivalent to minimizing the Lyapunov drift \eqref{eq:Lyapunov_drift} with the Lyapunov function defined in \eqref{eq:Lyapunov_function}. Further, constants $B$ and $\epsilon$ can be found to obtain the upper bound of the resulting Lyapunov drift. Therefore, the MP policy is concluded to stabilize the queuing network~\cite{varaiya_Max_2013}.
Nevertheless, the proof relies on several assumptions. A critical one is the unlimited queue capacity, which tends to violate in dense networks with short block lengths. 
In such scenarios, the MP controller potentially induces queue spillbacks or even gridlock \cite{gregoire_CapacityAware_2015}.
Additionally, the MP policy may also cause certain phases to be permanently inactivated~\cite{ji_DelayBased_2013}.
These issues have motivated the MP extensions discussed in Section~\ref{sec:background}, as well as the current study.

\section{Coordinated Maximum Pressure-plus-Penalty Control}
\label{sec:CMPP}

In this paper, we propose a coordinated adaptive traffic signal control policy, namely, Coordinated Maximum Pressure-plus-Penalty (CMPP), that takes advantage of the standard MP framework while overcoming the aforementioned issues by enforcing coordination among intersections. We further prove the strong stability still holds under CMPP control. 
In the remainder of this section, we first give a brief review of the Lyapunov optimization theorem~\cite{neely_Stochastic_2010}, the main theory used to establish the stability of CMPP (Section~\ref{sec:Lyaponov-min-drift-plus-penalty}), then describe the CMPP control policy and the corresponding penalty function (Section~\ref{sec:CMPP-penalty}), and finally outline the proof of stability (Section~\ref{sec:stability-proof}).

\subsection{Lyapunov Minimum Drift-Plus-Penalty Control}
\label{sec:Lyaponov-min-drift-plus-penalty}

The Lyapunov Minimum Drift-Plus-Penalty (LDPP) control is a prevalent control strategy in communication networks and queuing systems~\cite{neely_Stochastic_2010} with a primary goal to stabilize the queuing network while minimizing a penalty function over time. 
The general LDPP formulation is: 
\begin{equation}
\begin{aligned}\label{eq:min_avg_penalty}
    \min_\pi \quad & \bar{p}^\pi \\
    \textrm{s.t.} \quad & \boldQ^\pi(t) \textrm{ is stable,}
\end{aligned}
\end{equation}
where $\bar{p}^\pi = \underset{T \rightarrow \infty}{\lim} \frac{1}{T} \sum_{t = 0}^\top \bbE \{p^\pi(t)\}$ is the long-time average penalty, and $\boldQ^\pi(t)$ represents the queuing system under control policy $\pi$.
As an extension of the Lyapunov drift theorem, the Lyapunov Optimization theorem provides the condition of queuing stability with a bounded penalty:

\begin{theorem}[Lyapunov Optimization theorem]
\label{theorem:lyapunov_drift_plus_penalty}

Consider the Lyapunov function $L(\boldQ(t))$ defined in \eqref{eq:Lyapunov_function}. Suppose the initial queues satisfy $\bbE\{L(\boldQ(0))\} < \infty$, then the queuing system is strongly stable if there exist constants $B \geq 0, V > 0, \epsilon > 0$ and $p^*$ such that $\forall t$, 
\begin{align}\label{eq:LDPP-stability-condition}
\Delta(\boldQ(t)) \!+\! V \bbE\{p(t) \vert \boldQ(t)\} \leq B \!+\! V p^* \!-\! \epsilon \sum_{(l,m)\in\Lambda } q_{l,m}(t).
\end{align}
\end{theorem}

\begin{proof}
See the proof of Theorem 4.2 in \cite{neely_Stochastic_2010}.
\end{proof}

\subsection{CMPP Control and Penalty Function}
\label{sec:CMPP-penalty}

Let $\phi_i(t) =[\dots,\phi_{i,k}(t),\dots]^\top\in \{0,1\}^{K_i}$ denote the signal control of intersection $i$ and $\boldx_i(t) = [\phi_i(t)^\top, \dots,\phi_{i'}(t)^\top, \dots]^\top\in  \{0,1\}^{\sum_{j\in\setN_i\cup\{i\}} K_j}$ be the signal control of the neighborhood centered at intersection $i$, where $\setN_i$ denotes the set of neighboring intersections of $i$. 
At each time $t$, our proposed CMPP policy decides on the neighborhood signal control that maximizes the total pressure in the neighborhood minus a penalty $p_i(t)$ defined on the central intersection, which yields the control action as 
\begin{align} \label{eq:CMPP}
    \boldx^\text{CMPP}_i(t) = \underset{\boldx}{\arg \max}\left( \sum_{j\in\setN_i\cup \{i\}}\sum_{k\in\Phi_j} \gamma_{j,k}(t) \right) - V p_i(t),
\end{align}
where $V>0$ is a weight parameter. 

To address the issues of limited lane capacities and extensive green times, we design the penalty $p_{i,k}(t)$ to depend on the queue lengths at both intersection $i$ and its neighbors, as well as the elapsed time since the last time phase $k$ is activated. 
The resulting penalty function is given by
\begin{align} \label{eq:penalty_function_per_int}
\begin{split}
    p_i(t) = \sum_{(l,m)\in\Lambda_i}  \bigg (& \alpha^{(1)} h^{(1)}_{l,m}(t) + \alpha^{(2)}\sum_{p\in \setD_m} h^{(2)}_{l,m,p}(t) \\
    & + \alpha^{(3)} h^{(3)}_{l,m}(t) \bigg)
\end{split}
\end{align}
where $\alpha^{(k)}$ is the weight of penalty $h^{(k)}$, and the three penalty terms are specified as follows:
\begin{subequations}
\begin{align}
    h^{(1)}_{l,m}(t) &=
    \begin{cases}
        1, & \hat{q}_{l,m}(t+1) > \bar{q}, \\ 
        0, & \text{otherwise,}
    \end{cases} \label{eq:04_penalty_h1}\\
    h^{(2)}_{l,m,p}(t) &= 
    \begin{cases}
        1, & \hat{q}_{m,p}(t+1) > \bar{q}, \\
        0, & \text{otherwise,}
    \end{cases} \label{eq:04_penalty_h2}\\
    h^{(3)}_{l,m}(t) &= \sum_{k\in \setS_{l,m}} \phi_{i,k}(t) \left( \sum_{\tau = t-H}^t \phi_{i,k}(\tau) \right) \label{eq:04_penalty_h3}. 
\end{align}
\end{subequations}
In \eqref{eq:04_penalty_h1} and \eqref{eq:04_penalty_h2}, $\bar{q}$ denotes a threshold value for the queue length of each movement (e.g., lane capacity);
in \eqref{eq:04_penalty_h1}, $\hat{q}_{l,m}(t+1)$ is the predicted queue length as per \eqref{eq:store_and_forward_1};
in \eqref{eq:04_penalty_h2}, $\hat{q}_{m,p}(t+1)$ gives an upper bound on inflow vehicles to the downstream lane $p$ of $m$, i.e.,
\begin{align*}
\hat{q}_{m,p}(t+1) = q_{m,p}(t) - y_{m,p}(t) s_{m,p}(t) + y_{l,m}(t) s_{l,m}(t);
\end{align*}
and in \eqref{eq:04_penalty_h3}, $\setS_{l,m}$ is the set of phases that contains movement $(l,m)$ and $H$ specifies the backward tracing period. 
Hence, the first two penalty terms address lane capacities while the third penalizes continuous green time for certain phases.

\subsection{Stability Analysis}
\label{sec:stability-proof}

We finish this section by establishing the stability of the CMPP control policy. The proof is largely inspired by \cite{varaiya_Max_2013} while evoking Theorem \ref{theorem:lyapunov_drift_plus_penalty}. 
To start with, we introduce the following assumption, which is also used in \cite{varaiya_Max_2013}, to ensure the traffic demand does not exceed the control capability. 

\begin{assumption}[Bounded demand rate]
\label{assumption:demand_region}
The demand rate vector $\boldd=[\dots, d_{l,m},\dots]^\top$ is bounded such that there exists a control policy that satisfies
\begin{align}
    \bbE\{a_{l,m}(t)\} \leq \bbE\{b_{l,m}(t)\} + \epsilon \quad \forall (l,m) \in \Lambda,
\end{align}
for some $\epsilon>0$, where $a_{l,m}(t), b_{l,m}(t)$ are respectively total inflow and outflow of lane $l$ specified as 
\begin{subequations}
\begin{align}
    a_{l,m} &= \sum_{k \in \setU_l} (y_{k,l}(t)s_{k,l}(t) + d_l(t)) r_{l,m}(t), \label{eq:total_inflow_lane}\\
    b_{l,m} &= y_{l,m}(t) s_{l,m}(t) \label{eq:total_outflow_lane}.
\end{align}
\end{subequations}
\end{assumption}

Now we are ready to present the main theoretical result of this paper. \\

\begin{theorem}[Stability of CMPP]
Suppose Assumption~\ref{assumption:demand_region} holds
and that the initial queues satisfy $\bbE\{L(\boldQ(0))\} < \infty$, then the CMPP with penalty specified in \eqref{eq:penalty_function_per_int} guarantees strong stability. Further, the long-term average total queue length is bounded. 
\end{theorem}

\begin{proof}
As per \eqref{eq:CMPP}, the control action for intersection $i$ is determined by the total pressure within the neighborhood. 
Hence, we redefine the Lyapunov function as follows:

\begin{align} \label{eq:extended_Lyapunov_function}
    L(\boldQ(t)) &\overset{\triangle}{=} \frac{1}{2} \sum_{i\in\setI}\sum_{j\in\setN_i\cup \{i\}} \sum_{(l,m)\in\Lambda_j} q_{l,m}(t)^2 \notag \\
    &= \frac{1}{2} \sum_{(l,m) \in \Lambda} \kappa_{l,m} q_{l,m}(t)^2,
\end{align}
where $\kappa_{l,m}$ is a positive integer indicating how many times $q_{l,m}$ for each movement $(l,m) \in \Lambda$ is counted in $L(\boldQ(t))$.
According to Theorem 4.2 in \cite{neely_Stochastic_2010}, the stability condition \eqref{eq:LDPP-stability-condition} also applies to this general form of Lyapunov function. Following a similar derivation in \cite{varaiya_Max_2013,neely_Stochastic_2010}, we derive the upper bound on the corresponding Lyapunov drift as 
\begin{align}
    \Delta(\boldQ(t)) &\leq B - \epsilon \sum_{(l,m) \in \Lambda} \kappa_{l,m} q_{l,m}(t) \notag \\
    & \leq B - \epsilon' \sum_{(l,m) \in \Lambda} q_{l,m}(t) \label{eq:proof_inequality_bound},
\end{align}
where $\epsilon' = \epsilon \max_{(l,m)\in\Lambda} \kappa_{l,m}$ scales the original $\epsilon$ in \Cref{assumption:demand_region} by the maximum copy of queues in $L(\boldQ(t))$, 
and $B$ is a constant that satisfies
\begin{align}\label{eq:bound-B}
    B \geq \frac{1}{2} \sum_{(l,m) \in \Lambda} \kappa_{l,m} \bbE \left\{ (a_{l,m}(t) - b_{l,m}(t))^2 \vert \boldQ(t) \right\},
\end{align}
where $a_{l,m}$ and $b_{l,m}$ are defined in \eqref{eq:total_inflow_lane},~\eqref{eq:total_outflow_lane}. Under Assumption~\ref{assumption:demand_region}, both terms are bounded and thus $B$ exists.

Note that the penalty defined in \eqref{eq:penalty_function_per_int} has an upper bound
\begin{align}\label{eq:max-penalty}
    p_\text{max} = M_\text{max}\left(\alpha^{(1)} + D_\text{max}\alpha^{(2)} + \alpha^{(3)}H\right),
\end{align}
where $M_\text{max}=\max_{i\in\setI} |\Lambda_i|$ is the maximum number of movements at each intersection, and $D_\text{max}=\max_{m\in\setL} |\setD_m|$ is the maximum number of downstream lanes for each lane. 


Combining \eqref{eq:bound-B} and \eqref{eq:max-penalty}, we finally obtain the upper bound on the drift-plus-penalty as
\begin{align}\label{eq:CMPP-drift-plus-penalty-bound}
    \Delta(\boldQ(t)) \!+\! V \bbE\{p(t) \vert \boldQ(t)\} \leq B \!+\! V p_\text{max} \!-\! \epsilon' \sum_{(l,m) \in \Lambda} q_{l,m}(t),
\end{align}
which yields the strong stability as per \Cref{theorem:lyapunov_drift_plus_penalty}. 


Following \cite{neely_Stochastic_2010}, we proceed to show the long-term average total queue length is also bounded. To this end, we sum \eqref{eq:CMPP-drift-plus-penalty-bound} over time $\tau=0,\dots,t-1$ and yield
\begin{align}
    \bbE\{L(\boldQ(t))\} &- \bbE\{L(\boldQ(0))\} + V\sum_{\tau = 0}^{t-1} \bbE\{p(t) \vert \boldQ(t) \} \nonumber \\
    &\leq Bt + V p_{\text{max}} t - \epsilon' \sum_{\tau = 0}^{t-1} \sum_{(l,m) \in \Lambda} q_{l,m}(t).
\end{align}
Rearranging the inequality and dropping $\bbE\{L(\boldQ(t))\} \geq 0,\; \frac{1}{t}\sum_{\tau = 0}^{t-1} \bbE\{p(t) \vert \boldQ\} \geq 0$, we arrive at
\begin{equation*}
    \frac{1}{t} \sum_{\tau = 0}^{t-1} \sum_{(l,m) \in \Lambda} \kappa_{l,m} q_{l,m}(t) \leq \frac{B + V p_\text{max}}{\epsilon'} + \frac{\bbE\{L(\boldQ(0))\}}{\epsilon' t},
\end{equation*}
Taking $t \rightarrow \infty$, we finally get the upper bound on the total queue length as
\begin{equation*}
    \lim_{t\to\infty}\frac{1}{t} \sum_{\tau = 0}^{t-1} \sum_{(l,m) \in \Lambda} q_{l,m}(t) \leq \frac{B + V p_{\text{max}}}{\epsilon'}.
\end{equation*}
This concludes the proof.
\end{proof}

\section{Solution Algorithm for CMPP}
\label{sec:algo}

The key difference between CMPP and MP is the scope of local problems. Specifically, CMPP not only solves each intersection's own control action but also those of its neighbors. 
Therefore, CMPP can no longer be solved independently at each interaction but requires coordination among neighboring intersections. 

For notation simplicity, we drop the time index $t$ in this section. Let $\boldz=[\dots,\phi_i^\top,\dots]\in \{0,1\}^{K}$ be the control actions of all intersections and define an incidence matrix $M_i\in\{0,1\}^{(\sum_{j\in \setN_i\cup\{i\}} K_j)\times K}$ that transfer $\boldz$ into $\boldz_i=M_i\boldz$ that corresponds to $\boldx_i$. 
Accordingly, the CMPP policy prescribed in \eqref{eq:CMPP} is equivalent to solving the following optimization problem:
\begin{subequations}\label{eq:dist_optm}
\begin{align}
    \max_\boldz \quad &\sum_{i \in \setI} f_i(\boldx_i) \\
    \textrm{s.t.} \quad & \boldx_i - \boldz_i = 0, \quad \forall i\in \setI, \label{eq:consensus_constraint}
\end{align}
\end{subequations}
where the local objective is defined as
\begin{align*}f_i(\boldx_i)=\left(\sum_{j\in\setN_i\cup\{i\}}\sum_{k\in\Phi_j}\gamma_{j,k}(t)\right) - Vp_i(t).
\end{align*}

Problem \eqref{eq:dist_optm} is a distributed optimization with consensus constraints \cite{yang_Survey_2019}, though the binary decision variables $\boldx_i$ and $\boldz$ bring particular challenges to the solution procedure. In what follows, we detail two consensus algorithms that solve \eqref{eq:dist_optm} efficiently at each time step. The first is based on the alternating direction method of multipliers (ADMM) \cite{boyd_Distributed_2010} and the other is a greedy heuristic with a majority vote.

\subsection{ADMM}\label{subsec:admm_details}
ADMM features a decomposition-coordination algorithm that achieves the global optimum by iteratively solving local sub-problems \cite{boyd_Distributed_2010}. 
Although developed for convex optimization with continuous variables, ADMM shows a satisfactory performance in solving \eqref{eq:dist_optm} in our experiments. 

To obtain ADMM-based updating rules of each intersection, we first construct the augmented Lagrangian of \eqref{eq:dist_optm}
\begin{equation*}
    L(\boldx, \boldz, \boldlambda) = \sum_{i \in \setI} \left( f_i(\boldx_i) - \boldlambda_i^\top (\boldx_i \!-\! \boldz_i) - \frac{\rho}{2} \left| \left|\boldx_i \!-\! \boldz_i \right| \right|_2^2 \right),
\end{equation*}
where $\boldlambda_i$ is the dual variable introduced for each consensus constraint \eqref{eq:consensus_constraint}, and $\rho>0$ is a penalty parameter. We then derive the following iteration rules that are executed by each intersection independently. Namely, in each iteration, each intersection $i\in\setI$ performs
\begin{subequations}
\begin{align}
    \boldx_i^{k+1} &= \underset{\boldx_i}{\arg \max} f_i(\boldx_i) - \boldx^\top_i\boldlambda_i^k  - \frac{\rho}{2} \|\boldx_i \!-\! \boldz_i^k\|^2_2,   \label{eq:ADMM_x_update}\\
    (\boldz)_i^{k+1} &= \underset{(\boldz)_i}{\arg\max} \sum_{j\in\setN_i\cup\{i\}} \left((\boldlambda_j^k)_i + \rho (\boldx_j^{k+1})_i\right)^\top (\boldz)_i, \label{eq:ADMM_z_update}\\
    \boldlambda_i^{k+1} &= \boldlambda_i^k + \rho(\boldx_i^{k+1} - \boldz_i^{k+1}). \label{eq:ADMM_lambda_update}
\end{align}
\end{subequations}
In \eqref{eq:ADMM_z_update}, $(\boldz)_i\in\{0,1\}^{K_i}$, different from $\boldz_i$, denotes the subsequence in $\boldz$ that corresponds to the control of intersection $i$. The same notations apply to $(\boldlambda_j)_i, (\boldx_j)_i$. 

Note that \eqref{eq:ADMM_x_update} and \eqref{eq:ADMM_lambda_update} are directly derived from the general updating rules of ADMM, while \eqref{eq:ADMM_z_update} requires some additional decomposition, which is delineated in Appendix~\ref{appendix:ADMM_z_update}. 


\subsection{Greedy Heuristic}
Although ADMM can solve the subproblems efficiently in a distributed manner, it still requires quite a few iterations to reach a consensus, which may take extensive computation time in large networks. Hence, we develop another greedy algorithm with much lower computational complexity. 
The core idea is to perform a majority vote in the neighborhood of the intersection with the minimum local objective when a conflict emerges.
Accordingly, in each consensus iteration, at least one intersection determines its control action, which yields a computational complexity of $\mathcal{O}(|\setI|)$. 
The algorithm is detailed in \Cref{alg:greedy}.

\begin{algorithm}[t]
\caption{Greedy algorithm for CMPP} \label{alg:greedy}
\hspace*{\algorithmicindent} \textbf{Input:} $f_i(\boldx_i)$ 
\hspace*{2ex} \textbf{Output:} $\boldx^*_i$
\begin{algorithmic}[1]
    \State \textbf{Initialize}: Set $\setI_\text{DET}=\emptyset$.
    \While{$|\setI_\text{DET}|< N $}
    \State $\forall i\in \setI\setminus\setI_\text{DET}$, locally solve optimal control actions 
    \begin{equation*}
        \boldx^*_i = \underset{\boldx}{\arg \max} f_i(\boldx),
    \end{equation*}
    given $\boldx_j^*,\forall j\in \setI_\text{DET}$.  
    \State Compute local objectives $f_i^*=f_i(\boldx_i^*),\;\forall i\in \setI$.

    \State $\forall i\in \setI\setminus\setI_\text{DET}$,
    \If{the consensus conditions $(\boldx_i)_i = (\boldx_j)_i, \; 
    (\boldx_i)_j = (\boldx_j)_j, \; \forall j \in \setN_i$ hold}
    \State Add $i$ and $j,\forall j\in\setN_i$ to $\setI_\text{DET}$.
    \EndIf
    
    \State $\forall i\in \setI\setminus\setI_\text{DET}$,
    \If{$f_i^* < f_j^*,\;\forall j\in \setN_i$}
    \State Determine $\boldx_i^*$ by majority vote of neighbors.
    \State Add $i$ to $\setI_\text{DET}$.
    \EndIf
    
    \EndWhile
\end{algorithmic}
\end{algorithm}

Similar to ADMM, the greedy algorithm can be implemented in a distributed fashion. Specifically, Lines 5-8 and Lines 9-13 can be conducted in parallel, where the addition of intersections in Lines 7 and 12 are first performed on local copies of $\setI_\text{DET}$. These updates are then merged to update the global $\setI_\text{DET}$. 
Although the greedy algorithm lacks an optimality guarantee, it demonstrates satisfactory performance in simulations, as will be shown in Section~\ref{sec:sim}.

\section{Simulation Experiments}\label{sec:sim}

We evaluate the performance of our proposed CMPP controller against several benchmarks using CityFlow, an open-source traffic simulator for large-scale signal control~\cite{zhang_CityFlow_2019}.

\subsection{Simulation Environment}

\begin{figure*}[htbp]
    \centering
    \begin{subfigure}[t]{0.5\textwidth}
        \centering
        \includegraphics[height=2in]{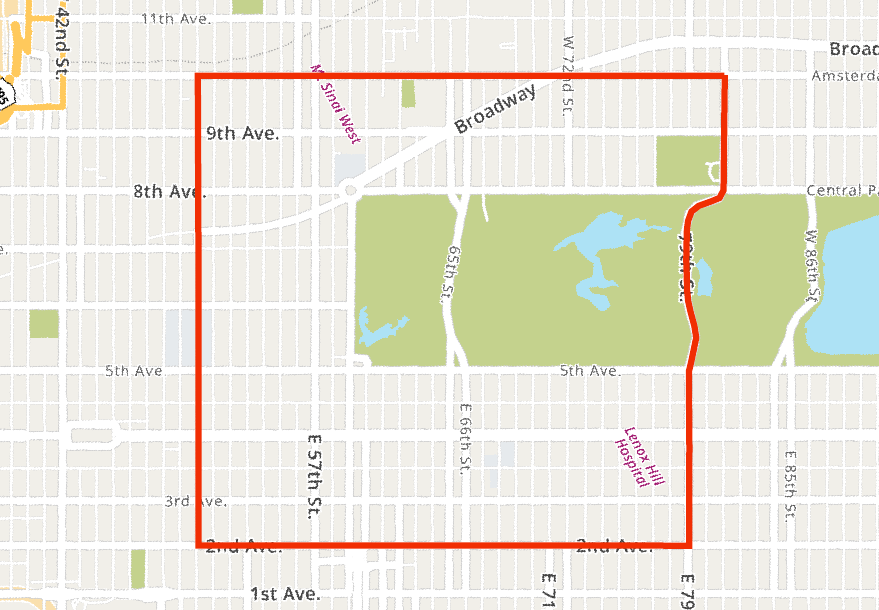}
        \caption{Region of interest.}
        \label{fig:road_network_map}
    \end{subfigure}%
    \begin{subfigure}[t]{0.5\textwidth}
        \centering
        \includegraphics[height=2in]{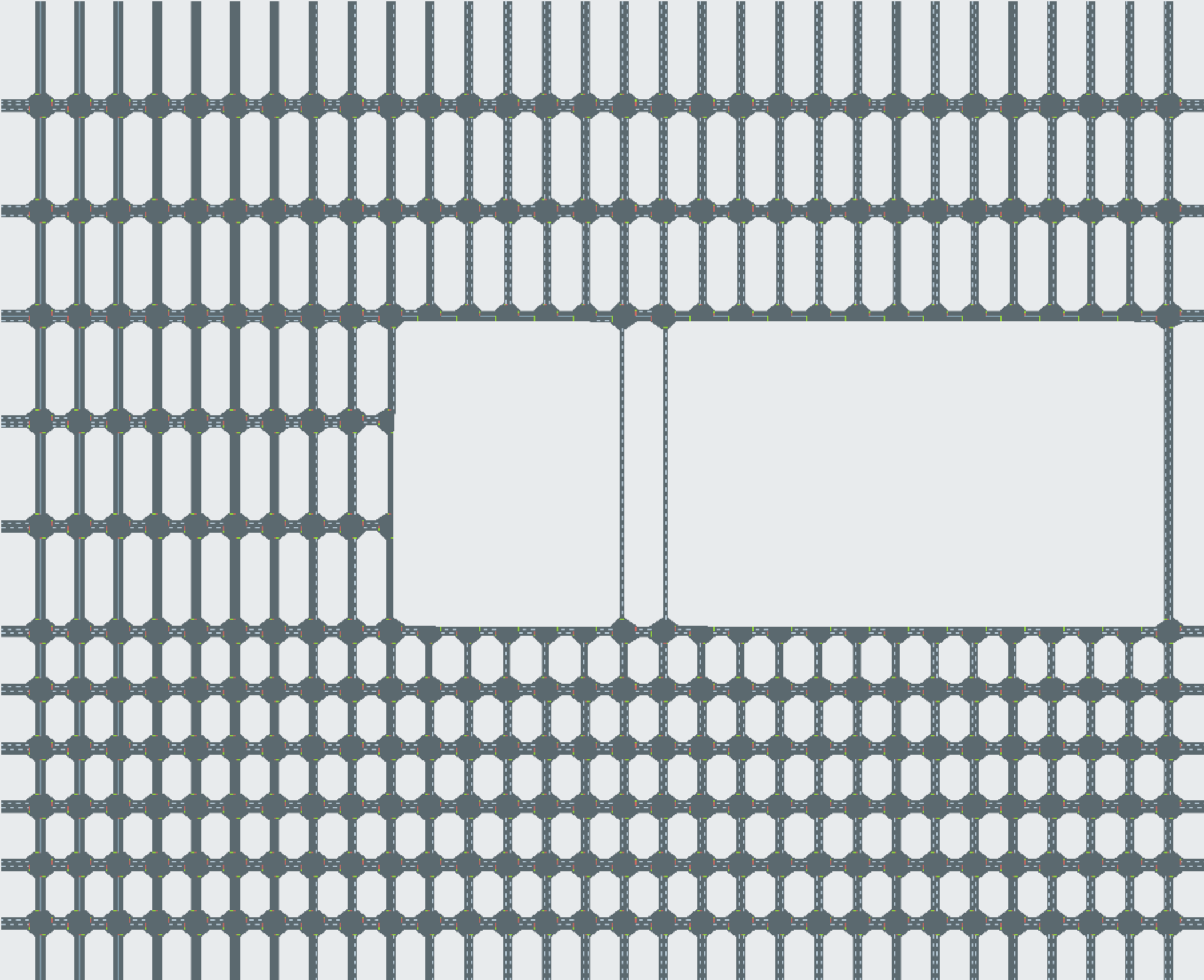}
        \caption{Road network in the simulator.}
        \label{fig:road_network_sim}
    \end{subfigure}
    \caption{Simulated traffic network based on Midtown Manhattan.}
    \label{fig:road_network}
\end{figure*}

\begin{figure}[htbp]
    \centering
    \includegraphics{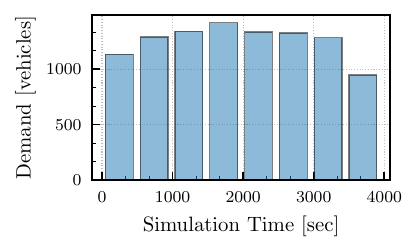}
    \caption{Traffic demand over simulation horizon.}
    \label{fig:demand}
\end{figure}

The simulation scenario is constructed based on the road network and traffic patterns in the midtown area of Manhattan, see \Cref{fig:road_network}. 
In total, the network has 290 intersections and features an average demand of \SI{9600}{veh \per \hour}. 
Each simulation is run for 4000 sec and the temporal demand pattern is illustrated in \Cref{fig:demand}. 
The free-flow travel speed is set to $\SI{30}{\km\per\hour}$ and the signal is updated every 20 sec.


\subsection{Benchmarks and Evaluation Metrics}

Besides CMPP, we implement three benchmark controllers: i) fixed-time (FT), where the signal plan is predefined; ii) classic MP; and iii) capacity-aware backpressure (CA-BP) proposed in \cite{gregoire_CapacityAware_2015} that largely resembles MP but accounts for the effect of limited queue capacity.
The results of CMPP using the above two solution algorithms are referred to as CMPP-ADMM and CMPP-Greedy, respectively. 

Due to the limit of space, we only report key performance metrics in this paper, including i) the average vehicle travel time, ii) the average vehicle waiting time (with speed less than \SI{0.1}{\m\per\sec}), iii) the number of vehicles traveling in the network at each time step, and iv) the average computation time for the control action at each time step.

\subsection{Main Results}\label{subsec:perf_evaluation}


This section reports the main simulation results of CMPP and benchmark controllers. After extensive parameter tuning, we use the penalty weights $\alpha^{(1)} = 4, \alpha^{(2)} = 2, \alpha^{(3)} = 0.1$ and the history horizon $H=3$. Since the weight for each penalty component is specified, $V$ is simply set to 1. 

\subsubsection{Average Travel Time}

\begin{figure}[htbp]
    \centering
    \includegraphics{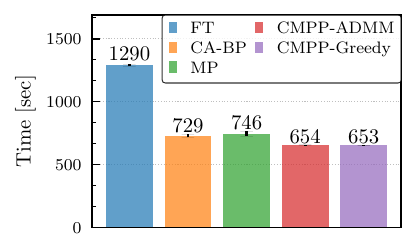}
    \caption{Average vehicle travel time [sec].}
    \label{fig:travel_time}
\end{figure}

\begin{figure}[htbp]
    \centering
    \includegraphics{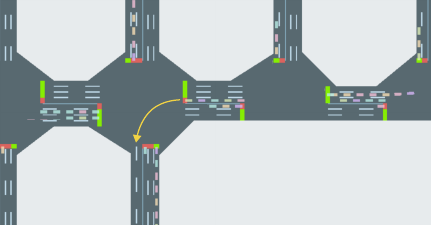}
    \caption{Example of queue spillover under MP.}
    \label{fig:MP_Congestion}
\end{figure}

\Cref{fig:travel_time} presents the average vehicle travel time. As expected, the FT controller performs the worst because it fails to adapt to the varying traffic dynamics. All other adaptive controllers achieve at least 40\% improvement compared to the FT baseline, while the MP controller leads to the longest average vehicle travel time among the four. Since the Manhattan network is quite dense with particularly short blocks along the avenues (the horizontal roads in \Cref{fig:road_network_sim}), queue spillovers are observed frequently in the movements from horizontal roads to vertical roads; see an example illustrated in \Cref{fig:MP_Congestion}. 

The spillover issue is partially addressed by CA-BP thanks to its specific attention to lane capacity, which ultimately results in slightly better performance than MP. Yet, CA-BP does not coordinate among neighboring intersections and thus tends to produce suboptimal control. The proposed CMPP control, on the other hand, outperforms all the benchmarks regardless of its solution algorithms. The additional 12\% saving from the MP controller is largely due to the coordination across intersections, which effectively prevents spillovers at high demand levels. 

\subsubsection{Average Waiting Time}

\begin{figure}[htbp]
    \centering
    \includegraphics{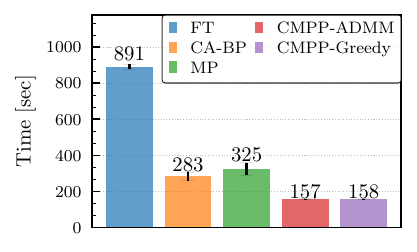}
    \caption{Average vehicle waiting time [sec].}
    \label{fig:waiting_time_per_vehicle}
\end{figure}

A more significant difference can be observed in \Cref{fig:waiting_time_per_vehicle}, which compares the average vehicle waiting time among the tested controllers. 
In addition to the expected extensive waits under FT, MP also results in quite a long waiting time, followed by CA-BP. This phenomenon is likely attributed to the second issue of MP discussed in Section~\ref{sec:MP}: since pressures at the current time step are the only metrics used to determine the phase activation, some phases may endure a long red time. 

\subsubsection{Network Congestion}

\begin{figure}[htbp]
    \centering
    \includegraphics{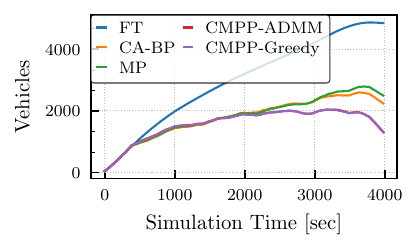}
    \caption{The number of vehicles in the network.}
    \label{fig:num_vehicles_in_network}
\end{figure}

\Cref{fig:num_vehicles_in_network} depicts the number of vehicles traveling in the network over the simulation horizon. Since the demand is relatively stable (see \Cref{fig:demand}), it also reflects the congestion level. In other words, an effective traffic signal controller should be able to maintain the vehicle number below a certain threshold. 

Since the network is empty at the beginning of the simulation, all controllers present the same increase before \SI{500}{\sec}. Afterward, the curve of FT grows faster than the others, demonstrating more severe congestion in the network. Both MP and CA-BP closely match CMPP until \SI{2000}{\sec}, from which the two curves start to deviate and increase at different rates. 
In contrast, both CMPP controllers well control the congestion and stabilize the number of vehicles in the network around 2000 in the second half of the simulation.


\subsubsection{Computation Time}

\Cref{tab:controller_times} reports the average computation time of each time step. As expected, MP achieves the highest efficiency thanks to its simple computations. CA-BP takes a slightly longer time because of the more complex expression of pressure. 
The two CMPP algorithms show drastically different computational efficiencies. Since the ADMM algorithm normally requires up to 10 consensus iterations to converge, each update requires up to 90 sec. 
Yet, its computational efficiency is expected to further improve with more computational resources and better-tuned stopping criteria. Nevertheless, ADMM can hardly outperform MP and CA-BP given the consensus iterations. On the other hand, the greedy algorithm shows a comparable computational efficiency with MP. On average, it takes 1.5 sec to complete each control update. 

\begin{table}[htb]
    \centering
    \caption{Average computation time for each signal update.}
    \label{tab:controller_times}
    \begin{tabular}{lc}
        \toprule
        \textbf{Control Method} & \textbf{Time [sec]} \\
        \midrule
        Max Pressure & 0.3 \\
        Capacity-Aware BP & 1.8 \\
        CMPP (ADMM) & 90.2 \\
        CMPP (Greedy) & 1.5 \\
    \bottomrule
    \end{tabular}
\end{table}

\subsubsection{Consensus Mechanism}
\begin{figure}[htbp]
    \centering
    \includegraphics{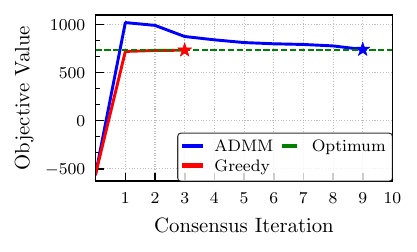}
    \caption{Example of consensus iterations under ADMM and Greedy algorithms.}
    \label{fig:consensus_iteration}
\end{figure}

We end this section by comparing the consensus mechanism of ADMM and greedy algorithms. As discussed above, while ADMM enforces consensus by directly handling consensus equality constraints (see \eqref{eq:consensus_constraint}), the greedy algorithm achieves almost the same performance as ADMM with much higher computational efficiency.  
As shown in \Cref{fig:consensus_iteration}, although both algorithms converge to the same global optimum, ADMM takes more iterations to converge and thus requires a longer computation time. 
On the other hand, the greedy algorithm is observed to often converge to the globally optimal solution even though it has no rigorous optimality guarantee. 
Hence, it is worthwhile to further explore whether the same result holds in general traffic networks.


\section{Conclusion}
\label{sec:conclusion}

This paper developed Coordinated Maximum Pressure-plus-Penalty (CMPP) control, a novel coordinated traffic signal control policy, and demonstrated its superior performance through extensive simulations. 
Inspired by Maximum Pressure (MP) control, CMPP adaptively updates traffic signals without predicting how traffic propagates over the network in future time steps and thus enjoys a high computational efficiency. 
Meanwhile, CMPP addresses existing issues of MP by enforcing coordination across intersections. In particular, it defines each local control problem over a neighborhood rather than a single intersection and introduces a penalty function that captures queue capacities and continuous green times. The resulting control policy was further proved to guarantee the queuing network stability by evoking the Lyapunov optimization theorem. 


To solve CMPP, we reformulated it as a distributed optimization problem and proposed two consensus algorithms. Our experiments show that CMPP outperforms benchmark controllers, regardless of its solution algorithm, in both individual vehicle travel and network congestion. 
The ADMM-based method usually requires a much longer computation time to converge. In contrast, our simulations show that the greedy heuristic achieves comparable computational efficiency as fully decentralized controllers (e.g., MP) without a considerable compromise of the control performance. 

Although this paper focuses on a signal specification of the penalty function, the CMPP control framework offers flexibility in designing the penalty to tackle various control scenarios and objectives. Yet, additional scrutiny is needed to extend the stability result for general penalty functions.
Additionally, more complex vehicle behaviors (e.g., dynamic rerouting) and signal coordination strategies (e.g., offset) can be integrated into future studies on CMPP.

\appendices

\section{ADMM update of $\boldz$} \label{appendix:ADMM_z_update}

The original ADMM iterative rule to update $\boldz$ is given by 
\begin{align}\label{eq:ADMM_z_update_original}
    \boldz^{k+1} &= \underset{\boldz}{\arg \min} \sum_{i = 1}^N \bigg( \Big[\boldlambda_i^k\Big]^\top \boldz_i - \frac{\rho}{2} \|\boldx_i^{k+1} - \boldz_i \|^2_2 \bigg). 
\end{align}
Following \cite{boyd_Distributed_2010}, we rewrite $\boldz$ as a vector of $N$ elements, each of which corresponds to the control of one intersection. 
Let $N_i=|\setN_i|$ be the number of neighbors and $\setM(i,j)$ denote the $j$-th neighbor of intersection $i$. 
Accordingly, \eqref{eq:ADMM_z_update_original} is expanded as 
\begin{align*}
    \boldz^{k+1} & = \underset{\boldz}{\arg \min}  \sum_{i = 1}^N  
    \begin{bmatrix}
    (\boldlambda_i)_i, \dots, (\boldlambda_i)_{\setM(i,N_i)}
    \end{bmatrix}
    \begin{bmatrix}
        (\boldz^k)_i \\
        \vdots \\ 
        (\boldz^k)_{\setM(i,N_i)}
    \end{bmatrix} \notag \\
    & \quad - \frac{\rho}{2} \left\| 
    \begin{bmatrix}
        (\boldx_i^{k+1})_i \\
        (\boldx_i^{k+1})_{\setM(i,1)}\\
        \vdots \\ 
        (\boldx_i^{k+1})_{\setM(i,N_i)}
    \end{bmatrix}
    -
    \begin{bmatrix}
        (\boldz^k)_i \\
        (\boldz^k)_{\setM(i,1)}\\
        \vdots \\ 
        (\boldz^k)_{\setM(i,N_i)}
    \end{bmatrix}
    \right\|_2^2\nonumber\\
    & = \underset{\boldz}{\arg \min} \sum_{i = 1}^N  \sum_{j\in \setN_i\cup\{i\}} (\boldlambda_i^k)_j^\top (\boldz)_j \notag \\
    & \quad - \frac{\rho}{2}\left( \|(\boldx_i^{k+1})_j\|^2_2 - 2(\boldx_i^{k+1})_j^\top (\boldz)_j + \|(\boldz)_j\|^2_2\right) \nonumber\\
    & = \underset{\boldz}{\arg \min} \sum_{i = 1}^N  \sum_{j\in \setN_i\cup\{i\}} \left((\boldlambda_i^k)_j + \rho (\boldx_i^{k+1})_j\right)^\top (\boldz)_j. 
\end{align*}
This allows us to further decompose the update and distribute it to each intersection $i$ as follows:
\begin{align*}
    (\boldz)_i^{k+1}  = \underset{(\boldz)_i}{\arg \min} \sum_{j\in \setN_i\cup\{i\}} \left((\boldlambda_j^k)_i + \rho (\boldx_j^{k+1})_i\right)^\top (\boldz)_i. 
\end{align*}

\balance

\bibliographystyle{IEEEtran}
\bibliography{reference}

\end{document}